\documentclass[aps,preprint,epsfig,rotate]{revtex4}

\usepackage{graphicx}
\usepackage{bm}
\usepackage{epsfig}

\input epsf

\begin{document}
\title{On the $Q$-dependence of the lowest-order QED corrections and other properties of the ground $1^1S-$states in the two-electron ions}

 \author{Alexei M. Frolov}
 \email[E--mail address: ]{afrolov@uwo.ca}

\affiliation{Department of Applied Mathematics \\
 University of Western Ontario, London, Ontario N6H 5B7, Canada}

\date{\today}

\begin{abstract}

Formulas and expectation values which are need to determine the lowest-order QED corrections ($\sim \alpha^3$) and corresponding recoil (or finite mass) corrections
in the two-electron helium-like ions are presented. Other important properties of the two-electron ions are also determined to high accuracy, including the 
expectation values of the qausi-singular Vinti operator and $\langle r^{-2}_{eN} \rangle$ and $\langle r^{-2}_{ee} \rangle$ expectation values. Elastic scattering of 
fast electrons by the two-electron ions in the Born approximation is considered. Interpolation formulas are derived for the bound state properties of the two-electron 
ions as the function of the nuclear electric charge $Q$.  

\noindent 
PACS number(s): 31.10.+z and 31.15.-p and 31.90.+s

\noindent 
30.12.2014, Preprint-2015-7-1/18 (this is 1st version 07 July 2015) [at.phys.], 18 pages.

\end{abstract}

\maketitle
\newpage

\section{Introduction}

In this contribution we determine the lowest-order Quantum Electrodynamic corresctions and other bound state properties in some two-electron ions with relatively large 
nuclear charges $Q$, where $Q \le 28$. In general, the lowest-order QED corrections are relatively small $\sim \alpha^3$ for the ground and low-excited states in the 
two-electron ions. These QED corrections arise from interactions of the bound electrons in ions and virtual photons. General theory of the lowest-order QED corrections in 
atoms/ions was considered in \cite{BS}. Since the first edition of that book the problem of lowest-order QED corrections was re-considered a number of times in different 
books and textbooks (see, e.g., \cite{AB}, \cite{Grein}). Note that often the lowest-order QED correction is called the `Lamb shift',or even `Bethe logarithm'. In reality, 
such a correction always contains a number of terms and Bethe logarithm is only one of them. Two other terms are the electron-electron contact term and Araki-Sucher term 
which contains the expectation value of the singular operator $\langle \frac{1}{r^{3}_{ee}} \rangle$. The explicit formula for the lowest-order QED correction is given in 
the third Section III. The formula presented there is based on an assumption that the nuclear mass is infinite, but in actual atomic systems the mass of the central `heavy' 
nucleus is always finite. Therefore, it is important to have a formula which allows one to determine the first-order correction upon the finite nuclear mass. 

By using highly accurate wave functions constructed in this study we can evaluate a number of other bound state properties for these two-electron ions, including the $\langle 
r^{-2}_{eN} \rangle$ and $\langle r^{-2}_{ee} \rangle$ expectation values, expectation value of the Vinti quasi-singular operator $\frac{{\bf r}_{31} \cdot {\bf 
r}_{32}}{r^{3}_{31}}$ and $\langle r^{2 k}_{eN} \rangle$ expectation values for $k = 2, 4, 6, \ldots$ which are needed to evaluate angular dependence of the elastic scattering 
cross-sections of the fast electrons by the two-electron, helium-like ions in the Born approximation. We also consider an important problem of $Q-$dependence for a number of 
expectation values determined in this study. 
  
\section{Bound state wave functions}

Modern highly accurate calculations of the lowest-order QED corrections in few-electron atomic systems, including two-electron ions, are performed with the 
use of the non-relativistic wave functions which are determined as the solutions of the Schr\"{o}dinger equation \cite{FrFi} for the bound states $H \Psi = 
E \Psi$, where $E < 0$ and $H$ is the non-relativistic Hamiltonian of the two-electron ion with the nuclear electric charge $Q e$: 
\begin{equation}
 H = -\frac{\hbar^{2}}{2 m_{e}} \Bigl( \nabla^{2}_{1} + \nabla^{2}_{2} + \frac{m_e}{M_N} \nabla^{2}_{N} \Bigr) - \frac{Q e^2}{r_{32}} - \frac{Q e^2}{r_{31}} 
 + \frac{e^2}{r_{21}} \label{Ham}
\end{equation}
where $\nabla_{i} = \Bigl( \frac{\partial}{\partial x_{i}}, \frac{\partial}{\partial y_{i}}, \frac{\partial}{\partial z_{i}} \Bigr)$ and $i = 1, 2, 3 (= N)$, 
where the notation $N$ (also 3) stands for the central atomic nucleus. In this equation and below the notation $\hbar$ stands for the reduced Planck constant, i.e. 
$\hbar = \frac{h}{2 \pi}$, and $e$ is the elementary electric charge. Note also that Eq.(\ref{Ham}) and everywhere below in this study the particles 1 and 2 mean 
the electrons, while the particle 3 (or $N$) is the atomic nucleus with the mass $M_N \gg m_e$. In real atomic calculations such a mass can also be an infinite.  
For light atoms and ions it is very convenient to perform all bound state calculations in atomic units where $\hbar = 1, m_e = 1$ and $e = 1$. In these units the 
velocity of light in vacuum $c$ numerically coincides with the inverse value of the dimensionless fine structure constant, i.e. $c = \alpha^{-1}$, where $\alpha 
= \frac{e^2}{\hbar c} \approx$ 7.2973525698$\cdot 10^{-3}$ \cite{CRC}. In atomic units the Hamiltonian from Eq.(\ref{Ham}) is written in the form
\begin{equation}
 H = -\frac12 \Bigl( \nabla^{2}_{1} + \nabla^{2}_{2} + \frac{1}{M_N} \nabla^{2}_{N} \Bigr) - \frac{Q}{r_{32}} - \frac{Q}{r_{31}} + \frac{1}{r_{21}} \label{Ham1}
\end{equation}
where the nuclear mass $M_N$ must be expressed in the electron mass $m_e$.

Formally, in calculations of the lowest-order QED corrections one needs to use the truly relativistic wave functions determined from the corresponding Dirac equation 
for the bound atomic states. At this moment, however, highly accurate solutions of the Dirac equation can be derived for one-electron atoms and ions only. For 
two-electron atoms and ions the relativistic wave functions can be constructed only approximately and overall accuracy of such an approximation is not high. It follows
from the well known fact that the electron-electron repulsion presents a great difficulty for analytical consideration in the Dirac theory of two-electron ions/atoms. 
In actual computations such an accuracy is substantially smaller than analogous accuracy obtained with the use of variational wave functions which were determined as
highly accurate solutions of the non-relativistic Schr\"{o}dinger equation. In this study to determine some properties of the two-electron atom(s) and ions we also apply 
the non-relativistic wave functions. All wave functions used in our analysis have been obtained as the solutions of the non-relativistic Schr\"{o}dinger equation with 
the Hamiltonian, Eq.(\ref{Ham1}). Note that this approach with the non-relativistic wave functions works, if (and only if) we can derive the non-relativistic variational 
wave functions of very high accuracy. Indeed, all expectation values below are determined in respect with the rules and formulas of the perturbation theory. In such 
calculations it is extremely important to have extremely accurate non-relativistic wave functions constructed in zero-order approximation.

In this study highly accurate wave functions of the ground $1^{1}S(L = 0)-$states of the two-electron ions/atoms are constructed in the form of the following variational 
expansion (see, e.g., \cite{Fro98} and \cite{Fro2001})
\begin{eqnarray}
 \Psi &=& \Bigl( 1 + \hat{P}_{12} \Bigr) \sum_{i=1}^{N} C_{i} \exp(-\alpha_{i} r_{32} - \beta_{i} r_{31} - \gamma_{i} r_{21}) \label{exp1} \\
 &=& \sum_{i=1}^{N} C_{i} \Bigl[ \exp(-\alpha_{i} r_{32} - \beta_{i} r_{31} - \gamma_{i} r_{21}) + \exp(-\beta_{i} r_{32} - \alpha_{i} r_{31} - \gamma_{i} r_{21})
  \Bigr] \nonumber 
\end{eqnarray}
The expansion, Eq.(\ref{exp1}), is called the exponential variational expansion in the relative coordinates $r_{32}, r_{31}$ and $r_{21}$. Each of the three relative 
coordinates $r_{ij}$ is defined as the difference between the corresponding Cartesian coordinates of the two particles, e.g., $r_{ij} = \mid {\bf r}_i - {\bf r}_j \mid
= r_{ji}$. It follows from this definition that the relative coordinates $r_{32}, r_{31}$ and $r_{21}$ are translationally and rotationally invariant. The coefficients 
$C_i$ are the linear variational parameters of the expansion, Eq.(\ref{exp1}), while the parameters $\alpha_{i}, \beta_{i}$ and $\gamma_{i}$ are the non-linear 
parameters of this expansion. In general, the total energy of the ground $1^1S-$state of the two-electron ion depends upon the total number of basis functions $N$, 
Eq.(\ref{exp1}), used in calculations. The operator $\hat{P}_{12}$ in Eq.(\ref{exp1}) is the permutation operator for two identical particles (electrons). Very high 
efficiency of the variational expansion, Eq.(\ref{exp1}), in actual applications to the two-electron ions is related to the fact that all non-linear parameters 
$\alpha_{i}, \beta_{i}$ and $\gamma_{i}$ are carefully varied in such calculations. The overall accuracy of our variational expansion, Eq.(\ref{exp1}), for all two-electron 
ions mentioned in this study can be considered as very high and sufficient for all our current needs. For the neutral helium atom we have performed numerical calculations
with an extended numerical precision. This allows us to determine 34 - 35 stable decimal digits in the total energy. By performing these calculations for the helium atom we 
investigated numerical stability of our results for the helium atom and other two-electron ions mentioned in this study. It was found in these research that our `usual' wave 
functions which provide 23 - 25 stable decimal digits are sufficient for our purposes in this study.   
   
\section{Lowest-order QED corrections}

Numerical calculation of the lowest-order QED correction in few-electron ion/atom is the two-stage procedure. At the first stage one determines the lowest-order QED 
correction for the model ion/atom with the infinitely heavy nucleus, e.g., for the ${}^{\infty}$Ne$^{8+}$ and ${}^{\infty}$Al$^{11+}$ ions. Such a correction is 
designated below as $\Delta E^{QED}_{\infty}$. The explicit formula for the lowest-order QED correction $\Delta E^{QED}_{\infty}$ in the two-electron ion with infinitely 
heavy nucleus is written in the form (in atomic units) 
\begin{eqnarray}
 \Delta E^{QED}_{\infty} &=& \frac{8}{3} Q \alpha^3 \Bigl[ \frac{19}{30} - 2 \ln \alpha - \ln K_0 \Bigr] \langle \delta({\bf r}_{eN}) \rangle 
 + \alpha^3 \Bigl[ \frac{164}{15} + \frac{14}{3} \ln \alpha - \frac{10}{3} S(S + 1) \Bigr] \langle \delta({\bf r}_{ee}) \rangle \nonumber \\
 &-& \frac{7}{6 \pi} \alpha^3 \langle \frac{1}{r^{3}_{ee}} \rangle \label{eqf8}
\end{eqnarray}
where $\alpha = \frac{e^2}{\hbar c} = 7.2973525698 \cdot 10^{-3}$ is the fine structure constant (see above), $Q$ is the nuclear charge (in atomic units) 
and $S$ is the total electron spin. The ground states in all two-electron ions considered in this study are the singlet states with $S = 0$. Also, in this 
formula $\ln K_0$ is the Bethe logarithm (see, e.g., \cite{BS}, \cite{AB}). 

The structure of the three-term formula, Eq.(\ref{eqf8}), has the following physical sense. Interactions of the bound electrons in ions and virtual photons leads to the 
variations in the electron density distribution. To determine the lowest-order QED corrections in two-electron ions the crucial factors are the electron density 
distributions at the atomic nucleus and at the second electron. This expalins the appearence of the two contact terms in Eq.(\ref{eqf8}): electron-electron nucleus contact 
term ($\sim \langle \delta({\bf r}_{eN} \rangle$) and electron-electron contact term ($\sim \langle \delta({\bf r}_{ee} \rangle$). The last term in Eq.(\ref{eqf8}) is called 
the Araki-Sucher term, or Araki-Sucher correction, since this correction was obtained and investigated for the first time by Araki and Sucher \cite{Araki}, \cite{Such}. 
Formally, this term represents the spatial derivative of the electron density at the nearest vicinity of the electron-electron collision point. Note that the expectation 
value of the term $\langle \frac{1}{r^{3}_{ee}} \rangle$ is singular, i.e., it contains the regular part and non-zero divergent part which is also called the principal part 
of the expectation value. General theory of the singular exponential integrals was developed in our earlier works (see, e.g., \cite{Fro2007} and references therein). In 
particular, in \cite{Fro2007} we have shown that the $\langle \frac{1}{r^{3}_{ee}} \rangle$ expectation value is determined by the following formula
\begin{eqnarray}
  \langle \frac{1}{r^{3}_{ee}} \rangle = \langle \frac{1}{r^{3}_{ee}} \rangle_R + 4 \pi \langle \delta({\bf r}_{ee}) \rangle \label{eqf85}
\end{eqnarray} 
where $\langle \frac{1}{r^{3}_{ee}} \rangle_R$ is the regular part of this expectation value and $\langle \delta({\bf r}_{ee}) \rangle$ is the expectation value of the 
electron-electron delta-function. The presence of non-zero divergent (or singular) parts in singular expectation values directly follows from the fact that the 
corresponding operators are self-conjugate. It is interesting to note that `to be self-conjugate' is the most fundamental and important property of any `symmetric' operator in 
QED. All other properties of such operators can be considered as secondary. Here we cannot discuss this interesting, but non-trivial problem. Briefly, we can only say that the 
overall contribution of the principal (or singular) part of the $\frac{1}{r^{3}_{ee}}$ operator is reduced to the expectation value of the electron-electron delta-function with 
an additional `spherical' factor 4 $\pi$. Formally, in \cite{Fro2007} the equality, Eq.(\ref{eqf85}), was derived for the exponential variational expansion, Eq.(\ref{exp1}), only. 
However, it can be shown that the same equality is true in the general case. Analogous formula can be written for the electron-nucleus expectation value $\langle \frac{1}{r^{3}_{eN}} 
\rangle$ which arises below. 
 
For the actual two-electron ions with the finite nuclear masses we need to evaluate the corresponding recoil correction to the lowest-order QED correction, Eq.(\ref{eqf8}). In atomic 
units such a correction to the lowest-order QED correction is written in the form
\begin{eqnarray}
 \Delta E^{QED}_{M} &=& \Delta E^{QED}_{\infty} - \Bigl(\frac{2}{M} + \frac{1}{M + 1} \Bigr) \Delta E^{QED}_{\infty} 
 + \frac{4 Q^2 \alpha^3}{3 M} \Bigl[ \frac{31}{3} + \frac{2}{Q} - \ln \alpha - 4 \ln K_0 \Bigr] \langle \delta({\bf r}_{eN}) \rangle \nonumber \\
 &+& \frac{7 \alpha^3}{3 \pi M} \langle \frac{1}{r^{3}_{eN}} \rangle \label{eqf9}
\end{eqnarray} 
where $M \gg m_e$ is the nuclear mass. The difference $\Delta E^{QED}_{M} - \Delta E^{QED}_{\infty}$ is the recoil shift of the lowest order QED correction. In some earlier works the 
recoil shift was defined as the absolute value of this difference and expressed in $MegaHertz$, or $MHz$, for short (1 $a.u.$ = 6.579683920729$\cdot 10^9$ $MHz$). The inverse mass 
$\frac{1}{M}$ in Eq.(\ref{eqf9}) is a small parameter which is (in atomic units) smaller than $\le 2.5 \cdot 10^{-5}$ for all isotopes of the two-electron ions which are heavier than 
the nitrogen two-electron ion N$^{5+}$. The dimensionless ratio $R = \frac{\mid \Delta E^{QED}_{\infty} - \Delta E^{QED}_{M} \mid}{\Delta E^{QED}_{\infty}}$ is always small (very small) 
and can be evaluated as $R \approx \frac{1}{M} \approx 1 \cdot 10^{-6} - 1 \cdot 10^{-5} \ll 1$ for two-electron ions considered in this study. 

\section{Specific component of the isotopic shifts. Vinti operator.}

Numerical calculations of the isotope shifts in few-electron atoms are performed with the use of the formula (see, e.g., \cite{Sob})
\begin{equation}
  E_M = E_{\infty} + E^{nms}_M +  E^{sms}_M = \Bigl( 1 - \frac{m_e}{m_e + M} \Bigr) E_{\infty} + E^{sms}_M \; \; \; \label{iso1}
\end{equation}
where $E_M$ and $E_{\infty}$ are the total energies of the actual ion/atom with the nucleus, which has a finite nuclear mass ($M$), and model ion/atom with the infinitely heavy nucleus. 
In this equation the notations $E^{nms}_M$ and $E^{sms}_M$ stand for the non-specific and specific mass shifts and $m_e$ is the electron mass. Physical meaning of the $E^{nms}_M$ 
value is clear from Eq.(\ref{iso1}). The explicit formula for the specific mass shift $E^{sms}_M$ in the A-electron atom/ion takes the form
\begin{eqnarray}
  E^{sms}_M = - \frac{\mu}{M m_e} \langle \Psi_{\infty} \mid \sum_{i=1}^{A-1} \sum_{j=2 (J > i)}^{A} \nabla_i \cdot \nabla_j \mid \Psi_{\infty} \rangle \; \; \; \label{iso2}
\end{eqnarray}
where $\mu = \frac{m_e M}{M + m_e}$ is the reduced electron mass and $M$ is the nuclear mass, while $m_e$ is the electron mass. The index $\infty$ designates the wave function
which correspond to the $A-$electron atom/ion with the infinitely heavy nucleus. In atomic units the formula, Eq.(\ref{iso2}), is written in the form
\begin{eqnarray}
  E^{sms}_M = - \frac{1}{M + 1} \langle \Psi_{\infty} \mid \sum_{i=1}^{A-1} \sum_{j=2 (J > i)}^{A} \nabla_i \cdot \nabla_j \mid \Psi_{\infty} \rangle \; \; \; \label{iso3}
\end{eqnarray}
Below we replace the notation $\Psi_{\infty}$ for the wave function by the traditional notation $\Psi$. This simplifies many formulas which follow. 

For two-electron atoms/ions considered in this study the right-hand side of Eq.(\ref{iso3}) is reduced to one expectation value $\langle \Psi \mid \nabla_1 \cdot \nabla_2 \mid 
\Psi \rangle = \langle \Psi \mid {\bf p}_1 \cdot {\bf p}_2 \mid \Psi \rangle$, where ${\bf p}_k = -\imath \nabla_k$ is the electron momentum of the $k-$th electron. The expression for 
this expectation value can be expressed in a number of ways, e.g., $\langle \nabla_1 \cdot \nabla_2 \rangle = - 2 \frac12 \langle \nabla^2_e \rangle + \frac12 \langle \nabla^2_N \rangle$, 
where $-\frac12 \langle \nabla^2_e \rangle$ essentially coincides with the single-electron kinetic energy, while $-\frac12 \langle \nabla^2_N \rangle$ is the kinetic energy of the nucleus.   
This relations between the $\langle \Psi \mid {\bf p}_1 \cdot {\bf p}_2 \mid \Psi \rangle$ expectation value, single-electron kinetic energy and kinetic energy of the nucleus 
follows from the conservation of the momentum of the three-particle system, e.g., in the center-of-mass we can write ${\bf P}_{N} = {\bf p}_1 + {\bf p}_2$. It follows from this equation 
that $\langle P^2_{N} \rangle = 2 \langle p^2_e \rangle + 2 \langle {\bf p}_1 \cdot {\bf p}_2 \rangle$. 

However, in the middle of 1930's it was assumed that computation of the expectation value of any differential operator, e.g., ${\bf p}_1 \cdot {\bf p}_2$ and/or $p^2_e$ cannot give a 
numerically stable result. In that time people tried to express such `unstable' expectation values in terms of the linear combinations of other expectation values which are the functions of 
the electron-nucleus and electron-electron variables. In particular, the following formula for the $\langle {\bf p}_1 \cdot {\bf p}_2 \rangle$ expectation value has been derived by Vinti 
\cite{Vinti}
\begin{eqnarray}
     \langle {\bf p}_1 \cdot {\bf p}_2 \rangle = Q \langle \frac{{\bf r}_{31} \cdot {\bf r}_{32}}{r^{3}_{31}} \rangle + \frac12 \langle \frac{1}{r_{21}} \rangle \; \; \; \label{vin1} 
\end{eqnarray}
where $Q$ (or $Q e$) is the electric charge of the nucleus (particle 3), and particles 1 and 2 are the two electrons. Numerical calculations of the expectation values of the electron-electron  
repulsion $\langle r^{-1}_{21} \rangle = \langle r^{-1}_{ee} \rangle$ is straightforward. However, the first expectation value in the right-hand side of Eq.(\ref{vin1}) contains a singular
expression. Indeed, in the relative coordinates the expectation value $\langle ({\bf r}_{31} \cdot {\bf r}_{32}) / r^{3}_{31} \rangle$ is represented in the form 
\begin{eqnarray}
   \langle \frac{{\bf r}_{31} \cdot {\bf r}_{32}}{r^{3}_{31}} \rangle = \frac12 \langle \frac{1}{r_{31}} \rangle + \frac12 \Bigl( \langle \frac{r^2_{32}}{r^3_{31}} \rangle 
   - \langle \frac{r^2_{21}}{r^3_{31}} \rangle \Bigr) \; \; \; \label{vin2}
\end{eqnarray}
where the last two expressions in the right-hand side are singular, if they are considered separately of each other. For years this fact made impossible any highly accurate computation of the 
expectation values of the Vinti operator $\frac{{\bf r}_{31} \cdot {\bf r}_{32}}{r^{3}_{31}}$. Recent analysis, however, indicates that computations of this expectation value in the exponential 
expansion, Eq.(\ref{vin2}), is reduced to operations with the Frullanian integrals only. By the Frullanian (three-particle) integrals we mean three-particle integrals which look like singular, 
but, in reality, they are regular integrals \cite{Frul}, \cite{Jef}. The expectation value of the Vinti operator contains singularities in the second and third terms in the right-hand side, but 
these singularities cancell each other from the final expression which becomes regular (for more details, see, e.g., \cite{Fro2007}). Note that in atcual computations it is better to consider 
the expectation value of the quasi-singular operator $V_S = \frac12 \Bigl( \frac{r^2_{32}}{r^3_{31}} - \frac{r^2_{21}}{r^3_{31}} \Bigr)$, since the first term in the righ-hand side of 
Eq.(\ref{vin2}) is a regular expectation value which representes the Coulomb electron-electron repulsion. 

In this study we have calculated the expectation values of the $V_s$-operator for each of the two-electron 
ions considered in this study. These expectation values (in atomic units) can be found in Table III. It is interesting to investigate numerical changes of the $\langle \frac{{\bf r}_{31} \cdot 
{\bf r}_{32}}{r^{3}_{31}} \rangle$ and/or $\frac12 \Bigl( \langle \frac{r^2_{32}}{r^3_{31}} \rangle  - \langle \frac{r^2_{21}}{r^3_{31}} \rangle$ expectation values in the series of two-electron 
ions, i.e. when the nuclear electric charge $Q$ increases. By using our results from Table III one can solve this problem for the expectation values of the $V_s$ operator and after a few simple
recalculations for the expectation values of the Vitni operator.    

\section{Scattering of fast electrons by the two-electron ions in the Born approximation}

Another interesting problem which is of interest for the two-electron ions is the scattering of fast electrons at such ions. This problem is actual in physics of the hot atomic plasmas, which 
can be created in modern laboratory, used in various technical applications and observed in astrophysics. The problem is formulated as follows. As is well known the electric charge density distribution 
$\rho({\bf r})$ in arbitrary atom/ion can be represented in the form $\rho({\bf r}) = Q e \delta({\bf r}) - e n({\bf r})$, where $Q e$ is the electric charge of the central nucleus, $n({\bf r})$ 
is the electron density and $\delta({\bf r})$ is the electron-nucleus delta-function. The electron density is normalized by the condition $\int n({\bf r}) d^3{\bf r} = N_e$, where $N_e$ is the 
number of bounded electrons. For atoms we always have $N_e = Q$, while in positively charged ions we have $N_e < Q$. For two-electron ions we have $N_e = 2$. From here one finds the following expression 
for the Born amplitude of elastic scattering:
\begin{eqnarray}
 f_B(q) = f_B\Bigl(2 k \sin\frac{\theta}{2}\Bigr) = - \frac{2}{a_0 q^2} [ Q - {\cal F}_e(q)] = - \frac{2}{a_0 q^2} [ Q - \int \exp(-\imath {\bf q} \cdot {\bf r}) n({\bf r}) d^3{\bf r}] \; \; \;
 \label{born1}
\end{eqnarray}
where $a_0$ is the Bohr radius and the value ${\cal F}_e(q)$ is the electron form-factor of the ion (see, e.g., \cite{Bet1964}). For two-electron ions with large $Q$ ($Q \ge 2$) the electron density 
is a spherically symmetric function, i.e. $n({\bf r}) = n(r)$. In these cases we have
\begin{eqnarray}
 {\cal F}_e(q) =  \int \exp(-\imath {\bf q} \cdot {\bf r}) n({\bf r}) d^3{\bf r} = 4 \pi \int_{0}^{+\infty} \frac{\sin(q r)}{q r} n(r) r^2 dr \; \; \; \label{born2}
\end{eqnarray}
For small values of the transfered momentum the kernel of the last integral in Eq.(\ref{born2}) is represented as a power series in $q r$. This leads to the following formula for $f_B(q)$
\begin{eqnarray}
 f_B(q) = f_B\Bigl(2 k \sin\frac{\theta}{2}\Bigr) = - \frac{2}{a_0 q^2} \Bigl[ (Q - N_e) + \frac{q^2 \langle r^2 \rangle}{3!} - \frac{q^4 \langle r^4 \rangle}{5!} + \frac{q^6 \langle r^6 \rangle}{7!}
  - \frac{q^8 \langle r^8 \rangle}{9!} + \ldots \; \; \; \label{born25}
\end{eqnarray}
For very compact two-electron ions with $Q \ge 2$ the series in the right-hand side of Eq.(\ref{born25}) converges very fast. The differential cross-section in the Born approximation can easily
be obtained from the formula, Eq.(\ref{born25}). Finally, we arive to the formula
\begin{eqnarray}
 \frac{d \sigma}{d \Omega} \sim \mid f_B(q) \mid^2 = \frac{4 (Q - N_e)^2}{a^2_0 q^4} [1 + \frac{q^2 \langle r^2 \rangle}{3! (Q - N_e)} - \frac{q^4 \langle r^4 \rangle}{5! (Q - N_e)} + 
  \ldots]^2 \; \; \; \label{born3}
\end{eqnarray}
where $Q > N_e$. The first term in this formula corresponds to the usual Rutherford scattering, while other terms in Eq.(\ref{born3}) represent the consequtive corrections for the electron scattering 
forward which is markedly anysotropic, since $q^2 = 4 k^2 \sin^2\frac{\theta}{2}$. Largest deviations from the Rutherford scattering can be found for the small angles (i.e. for $\theta \approx 1$ and 
$q R_a \gg 1$, where $R_a$ is the radius of the atom), where the cross-section of the scattering forward can exceed the cross-section of the Rutherford scattering. Note that application of these 
formulas to the two-electron H$^{-}$ ion is not apppropriate (it leads to an answer which is inaccurate and has no direct physical sense).  

\section{Calculations and formulas for the $Q^{-1}$ expansions}

In this Section we discuss a few important details of actual calculations of the bound state properties of two-electron ions considered above. First of all, let us note that the total number of bound 
state properties evaluated in this study to very high accuracy significantly larger (for each two-electron ion) than a few properties shown in Tables I - III. In particular, we determine the both 
electron-nucleus and electron-electron cusp values, i.e. the following ratios of the expectation values
\begin{eqnarray}
 \nu_{eN} = \frac{\langle \delta({\bf r}_{eN}) \frac{\partial}{\partial r_{eN}} \rangle}{\langle \delta({\bf r}_{eN}) \rangle} \label{eqf6}
\end{eqnarray}
in the case of the electron-nucleus cusp, and
\begin{eqnarray}
 \nu_{ee} = \frac{\langle \delta({\bf r}_{ee}) \frac{\partial}{\partial r_{ee}} \rangle}{\langle \delta({\bf r}_{ee}) \rangle} \label{eqf61}
\end{eqnarray}
for the electron-electron cusp. These two expectation values must coincide with the known values of these cusps, i.e., with the following numerical values (in atomic 
units)
\begin{eqnarray}
 \nu_{eN} = - Q e^2 \frac{m_e M_N}{m_e + M_N} = - Q \frac{1}{1 + \frac{1}{M_N}} = -Q \Bigr( 1 + M^{-1}_N \Bigl)^{-1} \; \; \; , \; \; \; \nu_{ee} = 0.5 
 \; \; \; \label{eqf62}
\end{eqnarray}
where $M_N = \frac{M_N}{m_e}$ is the nuclear mass which can be finite (real), or infinite for model atomic systems.  

The coincidence of these two expectation values, Eqs.(\ref{eqf6}) - (\ref{eqf61}), with the predicted values, Eqs.(\ref{eqf62}), is a very effective test for the variational wave functions in any 
Coulomb few-body system. In reality, this test is often used to show the correctness of our expectation values of the two-particle delta-functions determined with the use of variatonal expansion, 
Eq.(\ref{exp1}). For instance, for the Al$^{11+}$, Ca$^{18+}$ and Co$^{25+}$ ions the computed electron-nucleus cusp values are: $\nu_{eN}$ = -13.000000009571, $\nu_{eN}$ = -20.000000015543 and
$\nu_{eN}$ = -27.000000010287, respectively. Analogous electron-electron cusp values are $\nu_{ee}$ = 0.50000000545, $\nu_{ee}$ = 0.50000000652 and $\nu_{ee}$ = 0.50000000875, respectively. This 
indicates clearly that all expectation values of the two-particle delta-functions used in this study are highly accurate. 

Second, in our calculations of the lowest-order QED corrections we have used the $Q^{-1}$ expansion to evaluate the numerical values of the reduced Bethe logarithm, i.e. for the $\ln \Bigl(
\frac{K_0}{Q^2 Ry} \Bigr) = \ln k_0$ value:
\begin{eqnarray}
 \ln k_0 = \ln \Bigl( \frac{K_0}{Q^2 Ry} \Bigr) = C_0 + \frac{C_1}{Q} + \frac{C_2}{Q^2} + \frac{C_3}{Q^3} + \ldots + \frac{C_n}{Q^n} + \ldots \approx \sum^{N}_{k=0} \frac{C_k}{Q^k}  \label{Qm1}
\end{eqnarray}
In actual computations this series converges very fast, since the coefficients $C_1, C_2, \ldots$ are significantly smaller than the first coefficient $C_0$ (this is the main advantage to apply the 
reduced Bethe logarithm $\ln k_0$ in actual calculations). Therefore, we can restrict ourselves by a relatively small $N$ in Eq.(\ref{Qm1}). In this study we chose $N = 6$. By using the known values 
of the Bethe logarithm for the two-electron He atom ($\ln K_0$ = 4.2701621981) and H$^{-}$ ($\ln K_0$ = 2.993044153), Li$^{+}$, Be$^{2+}$, B$^{3+}$, C$^{4+}$ ions (the same values of $\ln K_0$ which 
are mentioned in the first reference from \cite{Fro2007}) we have found the following values of the $C_k$ coefficients in Eq.(\ref{Qm1}):
\begin{eqnarray}
  C_0 &=&  2.9841349233 \; \; \; , \; \; \; C_1 = -0.0124321314 \; \; \;  \; \; \;  C_2 =  0.0235421443 \; \; \; \label{coeff} \\
  C_3 &=& -0.0008950491 \; \; \; , \; \; \; C_4 =  0.0068528122 \; \; \; , \; \; \; C_5 = -0.0081585463 \nonumber 
\end{eqnarray} 
Now, by using these six cefficients we find for the Co$^{25+}$ ions $\ln k_0$ = 2.9836752649 $\approx$ 2.9836753 and $\ln K_0$ = 9.4214269146 $\approx$ 9.4214269. Numerical evaluation of Bethe 
logarithms for other two-electron ions is performed analogously. Note that the formula, Eq.(\ref{Qm1}), can be applied to the ground $1^1S-$states in the two-electron ions/atoms only. For the excited
states the series, Eq.(\ref{Qm1}), must contain an additional term.    

Third, numerical calculation of the expectation value of the Vitni operator $\langle \frac{{\bf r}_{31} \cdot {\bf r}_{32}}{r^{3}_{31}} \rangle$ is reduced to computations of the `singular' part
of this operator $V_s =  \frac12 \Bigl( \frac{r^2_{32}}{r^3_{31}} - \frac{r^2_{21}}{r^3_{31}} \Bigr)$ (the corresponding numerical values can be found in Table III) and electron-electron repulsion 
energy in atomic units $\langle \frac{1}{r_{21}} \rangle = \langle r^{-1}_{21} \rangle$. These expectation values can be found in Table IV which also contains the expectation values of 
single-electron kinetic energies $\langle -\frac12 \nabla^2_e \rangle$ and triple delta-functions $\langle \delta_{321} \rangle$ (all values expressed in atomic units). 

In order to represent the expectation values determined in this study (see Tables I - IV) we can derive relatively simple and convenient interpolation formulas (or asymptotic formulas) which represent 
the $Q-$deppendence (or $Q^{-1}-$deppendence) of different expectation values. Such interpolation formulas are of great interest in actual applications to few-body systems, since in reality it is  
difficult to use (and even keep) huge tables of highly accurate data. However, if we know the explicit formula for the $Q-$expansions of these expectation values, then we can operate with a restricted
number of the expansion coefficients determined separately for each expectation value. Let us consider the $\langle r^{-1}_{21} \rangle = \langle r^{-1}_{ee} \rangle$ expectation values determined for 
different two-electron ions. Numerical values of this expectation value for different ions can be found in Table IV. As follows from this Table the expectation value $\langle r^{-1}_{ee} \rangle$ 
increases with $Q$. Therefore, we can approximate these expectation values by using the following $Q^{-1}$-expansion:
\begin{eqnarray}
 \langle r^{-1}_{ee} \rangle = A_2 Q^2 + A_1 Q + A_0 + \frac{B_1}{Q} + \frac{B_2}{Q^2} + \frac{B_3}{Q^3} + \ldots + \frac{B_n}{Q^n} + \ldots \label{Qexp1}
\end{eqnarray}
where the first three terms represent the regular part of the Laurent expansion (or series), while all terms with the negative powers of $Q$ represent the principal part of the Laurent series. By using
our data for the $\langle r^{-1}_{ee} \rangle$ expectation values from Table IV we can estimate the first few coefficients $A_2, A_1, A_0, B_1, B_2, B_3, \ldots$ from a least-square fit to our calculated
expectation values of $\langle r^{-1}_{ee} \rangle$ up to $Q = 28$. The four different sets of such coefficients are shown in Table V. They correspond to the cases when the expansion Eq.(\ref{Qexp1}) 
contains 10, 12 and 14 terms. It is intereting to illustrate the power of the general theory which is behind the $Q^{-1}$-expansions (see, e.g., \cite{Eps}). As folows from the comparison of the first two 
columns in Table V our guess about the explicit form of Eq.(\ref{Qexp1}) is wrong. Indeedn the first coefficient $A_2$ in the first column is almost zero, while the secod coefficient is significantly 
larger (in $10^8$ times!). This means we have to assume that $A_2 = 0$ identically, or, in other words, the first actual term in Eq.(\ref{Qexp1}) must be $A_1 Q$. This was corrected in calculations performed 
for the columns 2 - 4 in Table V. Briefly, this means that one can make a mistake in Eq.(\ref{Qexp1}), but general theory will correct it. Briefly, we can say that the correct form of the $q^{-1}$ expansion 
for each expectation value will be restored in any case, if the original computational data are truly highly accurate.  

Interpolation (or asymptotic) formulas for other expectation values can be derived analogously. Note that in actual applications one finds three following situations: (1) expectation value decreases when 
the nuclear charge $Q$ grows, (2) expectation value increases when the nuclear charge $Q$ increases, and (3) expectation value is alomst constant during $Q$ variations. For instance, all $\langle r^{k}_{eN} 
\rangle$ and $\langle r^{k}_{ee} \rangle$ expectation values with positive $k$ represent the first group. The expectation values $\langle r^{k}_{eN} \rangle$ and $\langle r^{k}_{ee} \rangle$ with negative 
$k$ belong to the second group. The reduced Bethe logarithm mentioned above corresponds to the third group. The explicit formulas of the $Q-$expansions in each of these cases can be derived analogously. 
Furthermore, this procedure can be applied to each of the expectation values presented in Tables I - IV.           
   
\section{Conclusions}

We have investigated the $Q-$dependencies of different bound state properties of the ground $1^1S-$states in the seires of two-electron ions which includes all ions from the hydrogen two-electron ion up to 
the two-electron nickel ion. A substantial part of this series (ions from N$^{5+}$ to Ni$^{26+}$) is of great interest, since these ions were not considered in earlier highly accurate studies. In particular, 
the lowest-order QED corrections for each of these ions have been determined to high accuracy for the first time. For convenience some results of such calculations are shown in Table VI where all results are 
given in atomic units. We also determine the specific component of the isotopic shifts in these ions by using the expectation values of the quasi-singular (or Frullanian) Vinti operator. Scattering of fast 
electrons by the two-electron ions is briefly discussed in the Born approximation. A brief theory of accurate interpolation formulas for $Q-$dependencies of different bound state properties is discussed and 
illustrated with a few examples.

\newpage
 \begin{table}[tbp]
   \caption{The total non-relativistic energies $E$ and $\langle r^{-2}_{eN} \rangle$ and $\langle r^{-2}_{ee} \rangle$ expectation value 
            for a number of two-electron atoms/ions in their ground $1^1S-$states (in atomic units). All nuclear masses are assumed to be 
            infinite.}
     \begin{center}
     \scalebox{0.85}{%
     \begin{tabular}{| c | c | c | c | c|}
      \hline\hline
  ion/atom & $Q$ &             $E$                         &  $\langle r^{-2}_{eN} \rangle$ & $\langle r^{-2}_{ee} \rangle$ \\
     \hline\hline              
   H$^{-}$     & 1 &   -0.5277510165443771965925           & 1.1166628245254385975 & 0.1551041525624275657 \\      
   He          & 2 &  -2.90372437703411959831115924519440  & 6.0174088670242866935 & 1.4647709233190671890 \\  
   Li$^{+}$    & 3 &   -7.27991341266930596491875          & 14.927623721321274552 & 4.0822327875220922168 \\  
   Be$^{2+}$   & 4 & -13.65556623842358670208051           & 27.840105671230351970 & 8.0288017817568097761 \\
   B$^{3+}$    & 5 & -22.03097158024278154165469           & 44.753485397059351468 & 13.307121456650923770 \\
   C$^{4+}$    & 6 &  -32.40624660189853031055685          & 65.667312718097517695 & 19.918006872526419836 \\
             \hline 
   N$^{5+}$   & 7  &  -44.781445148772704645183           & 90.58139550837612183 & 27.86179483571140529 \\
   O$^{6+}$   & 8  &  -59.156595122757925558542           & 119.4956378746231599 & 37.13865011529274487 \\
   F$^{7+}$   & 9  &  -75.531712363959491104856           & 152.4099865906842945 & 47.74866291286534100 \\
  Ne$^{8+}$   & 10 &  -93.906806515037549421417           & 189.3244097370080602 & 59.69188677567529073 \\
  Na$^{9+}$   & 11 &  -114.28188377607272189582           & 230.2388870075436700 & 72.96835550416178458 \\
  Mg$^{10+}$  & 12 &  -136.65694831264692990427           & 275.1534048677168016 & 87.57809148852457738 \\
             \hline 
  Al$^{11+}$ & 13 & -161.03200302605835987252          & 324.0679539487669773 & 103.5211101497819791 \\
  Si$^{12+}$ & 14 & -187.40704999866292631487          & 376.9825275594010858 & 120.7974224560008489 \\
  P$^{13+}$  & 15 & -215.78209076353716023462          & 433.8971207930373478 & 139.4070364215928324 \\
  S$^{14+}$  & 16 & -246.15712647425473932009          & 494.8117299699472423 & 159.3499580389784688 \\
  Cl$^{15+}$ & 17 & -278.53215801540009570337          & 559.7263522763435335 & 180.6261918784910823 \\
  Ar$^{16+}$ & 18 & -312.90718607661114879880          & 628.6409855238003586 & 203.2357414866408062 \\
             \hline 
  K$^{17+}$  & 19 & -349.28221120345316700447          & 701.5556279846605630 & 227.1786096576231830 \\
  Ca$^{18+}$ & 20 & -387.65723383315855621790          & 778.4702782768293685 & 252.4547986227706891 \\
  Sc$^{19+}$ & 21 & -428.03225432023469116264          & 859.3849352814886391 & 279.0643101854956014 \\
  Ti$^{20+}$ & 22 & -470.40727295513838395930          & 944.2995980832556961 & 307.0071458191845422 \\
  V$^{21+}$  & 23 & -514.78228997811177388135          & 1033.214265925954510 & 336.2833067393932757 \\
  Cr$^{22+}$ & 24 & -561.15730558958127234352          & 1126.128938179445216 & 366.8927939578833437 \\
              \hline
  Mn$^{23+}$ & 25 & -609.53231995807574620568          & 1223.043614314415181 & 398.8356083236148089 \\
  Fe$^{24+}$ & 26 & -659.90733322632780520901          & 1323.958293882988089 & 432.1117505542265143 \\
  Co$^{25+}$ & 27 & -712.28234551602655145614          & 1428.872976503642391 & 466.7212212604835555 \\ 
  Ni$^{26+}$ & 28 & -766.65735693155709991040          & 1537.787661849361804 & 502.6640209654591693 \\
     \hline\hline
  \end{tabular}}
  \end{center}
  \end{table}
\newpage
 \begin{table}[tbp]
   \caption{The $\langle \delta({\bf r}_{eN}) \rangle, \langle \frac{1}{r^{3}_{eN}} \rangle_R, \langle \delta({\bf r}_{ee}) \rangle$ and 
            $\langle \frac{1}{r^{3}_{ee}} \rangle_R$ expectation values for a number of two-electron atoms/ions in their ground $1^1S-$states 
            (in atomic units). The subscript $R$ is used below to designate the regular part of the expectation value. All nuclear masses are 
            assumed to be infinite.}
     \begin{center}
     \scalebox{0.85}{%
     \begin{tabular}{| l | l | l | l | l | l |}
      \hline\hline
  ion/atom & $Q$ & $\langle \delta({\bf r}_{eN}) \rangle$ & $\langle \frac{1}{r^{3}_{eN}} \rangle_R$ & $\langle \delta({\bf r}_{ee}) \rangle$ & $\langle \frac{1}{r^{3}_{ee}} \rangle_R$ \\
     \hline\hline              
   H$^{-}$     & 1 & 0.16455287284772 & -3.435594850575 & 0.0027379921262 &  0.064307887281 \\      
   He          & 2 & 1.81042931850139 & -53.67642660259 & 0.106345370634  & -0.347101795446 \\  
   Li$^{+}$    & 3 & 6.85200943432227 & -238.7091644015 & 0.533722536561  & -6.528108292960 \\   
   Be$^{2+}$   & 4 & 17.1981725446016 & -662.2679562154 & 1.522895351487  & -26.72596510717 \\
   B$^{3+}$    & 5 & 34.7587436609116 & -1437.181953992 & 3.312442112829  & -70.59566341481 \\
   C$^{4+}$    & 6 & 61.4435780565873 & -2682.795781398 & 6.141043971072  & -148.7264629873 \\
             \hline 
   N$^{5+}$   & 7  & 99.1625346267485 & -4523.555532659 & 10.247411460455 & -272.4138789959 \\ 
   O$^{6+}$   & 8  & 149.825472814519 & -7088.104852970 & 15.870266106959 & -453.5227647586 \\
   F$^{7+}$   & 9  & 215.342252155782 & -10508.65528876 & 23.248334282439 & -704.3954157123 \\ 
  Ne$^{8+}$   & 10 & 297.622732139792 & -14920.52185328 & 32.620344790203 & -1037.785453519 \\
  Na$^{9+}$   & 11 & 398.576772218307 & -20461.76606779 & 44.225027751185 & -1466.807903211 \\
  Mg$^{10+}$  & 12 & 520.114231810706 & -27272.91291287 & 58.301114053422 & -2004.900136214 \\
             \hline 
  Al$^{11+}$ & 13 & 664.144970311620 & -35496.72082712 & 75.087335056216 & -2665.790465008 \\
  Si$^{12+}$ & 14 & 832.578847097505 & -45277.99108426 & 94.822422421253 & -3463.472333880 \\
  P$^{13+}$  & 15 & 1027.32572153134 & -56763.40721339 & 117.74510801161 & -4412.182728211 \\
  S$^{14+}$  & 16 & 1250.29545296610 & -70101.39786688 & 144.09412382866 & -5526.383843605 \\
  Cl$^{15+}$ & 17 & 1503.39790074724 & -85442.01834226 & 174.10820197134 & -6820.747326894 \\
  Ar$^{16+}$ & 18 & 1788.54292421458 & -102936.8471892 & 208.02607460887 & -8310.140582248 \\
             \hline 
  K$^{17+}$  & 19 & 2107.64038270360 & -122738.8951897 & 246.08647396232 & -10009.61476079 \\
  Ca$^{18+}$ & 20 & 2462.60013554641 & -145002.5246139 & 288.52813229153 & -11934.39414078 \\
  Sc$^{19+}$ & 21 & 2855.33204207246 & -169883.3771019 & 335.58978188593 & -14099.86666982 \\
  Ti$^{20+}$ & 22 & 3287.74596160910 & -197538.3088595 & 387.51015505776 & -16521.57548801 \\
  V$^{21+}$  & 23 & 3761.75175348198 & -228125.3321072 & 444.52798413712 & -19215.21128676 \\
  Cr$^{22+}$ & 24 & 4279.25927701537 & -261803.5619184 & 506.88200146822 & -22196.60538551 \\
              \hline
  Mn$^{23+}$ & 25 & 4842.17839153237 & -298733.1677353 & 574.81093940663 & -25481.72342952 \\
  Fe$^{24+}$ & 26 & 5452.41895635514 & -339075.3289712 & 648.55353031710 & -29086.65962870 \\
  Co$^{25+}$ & 27 & 6111.89083080505 & -382992.1942021 & 728.34850657198 & -33027.63147071 \\
  Ni$^{26+}$ & 28 & 6822.50387420288 & -430646.8435300 & 814.43460055003 & -37320.97485193 \\ 
     \hline\hline
  \end{tabular}}
  \end{center}
  \end{table}
\newpage
 \begin{table}[tbp]
   \caption{The $\langle r^{2}_{eN} \rangle, \langle r^{4}_{eN} \rangle$ and $V_s = \frac12 \Bigl( \langle \frac{r^2_{32}}{r^3_{31}} \rangle 
            - \langle \frac{r^2_{21}}{r^3_{31}} \rangle$ expectation values of the different two-electron atoms/ions in their ground 
            $1^1S-$states (in atomic units). All nuclear masses are assumed to be infinite.}
     \begin{center}
     \scalebox{0.85}{%
     \begin{tabular}{| l | l | l | l | l |}
      \hline\hline
  ion/atom & $Q$ & $\langle r^{2}_{eN} \rangle$ & $\langle r^{4}_{eN} \rangle$ & $\frac12 \Bigl( \langle \frac{r^2_{32}}{r^3_{31}} \rangle  - \langle \frac{r^2_{21}}{r^3_{31}} \rangle \Bigr)$ \\
     \hline\hline              
   H$^{-}$     & 1 & 11.913699678051262                & 645.14454241221937                & -0.4642618530805959 \\      
   He          & 2 & 1.1934829950189353                & 3.9735649316629101                & -1.0010782750156071 \\ 
   Li$^{+}$    & 3 & 0.44627901120132874               & 0.52960197172324301               & -1.5089235297645549 \\
   Be$^{2+}$   & 4 & 0.23206731553080959               & 0.14056152936071842               & -2.0126039740505242 \\
   B$^{3+}$    & 5 & 0.14196918178327382               & 5.2083409587888660$\cdot 10^{-2}$ & -2.5147510387558219 \\
   C$^{4+}$    & 6 & 9.5739522935058123$\cdot 10^{-2}$ & 2.3539482331106725$\cdot 10^{-2}$ & -3.0161594174263226 \\
             \hline 
   N$^{5+}$   & 7  & 6.8904525396724106$\cdot 10^{-2}$ & 1.2141346213809833$\cdot 10^{-2}$ & -3.5171547716283449 \\
   O$^{6+}$   & 8  & 5.1955356759041028$\cdot 10^{-2}$ & 6.8816453528063574$\cdot 10^{-3}$ & -4.0178956847309996 \\
   F$^{7+}$   & 9  & 4.0570590003835947$\cdot 10^{-2}$ & 4.1863570102541125$\cdot 10^{-3}$ & -4.5184687145246674 \\
  Ne$^{8+}$   & 10 & 3.2556160988739701$\cdot 10^{-2}$ & 2.6907921048456992$\cdot 10^{-3}$ & -5.0189251377057326 \\
  Na$^{9+}$   & 11 & 2.6701867132630897$\cdot 10^{-2}$ & 1.8073903567941317$\cdot 10^{-3}$ & -5.5192972710023712 \\
  Mg$^{10+}$  & 12 & 2.2295730542814468$\cdot 10^{-2}$ & 1.2585800135898979$\cdot 10^{-3}$ & -6.0196064958067383 \\
             \hline 
  Al$^{11+}$ & 13 & 1.889667256146306$\cdot 10^{-2}$ & 9.031562973287720$\cdot 10^{-4}$ & -6.5198675240121025 \\
  Si$^{12+}$ & 14 & 1.621960949319265$\cdot 10^{-2}$ & 6.648054655635712$\cdot 10^{-4}$ & -7.0200908109741794 \\
  P$^{13+}$  & 15 & 1.407363963658752$\cdot 10^{-2}$ & 5.001508016520978$\cdot 10^{-4}$ & -7.5202839913624110 \\
  S$^{14+}$  & 16 & 1.232702890357691$\cdot 10^{-2}$ & 3.834614539301480$\cdot 10^{-4}$ & -8.0204527704226906 \\
  Cl$^{15+}$ & 17 & 1.088649409859439$\cdot 10^{-2}$ & 2.989044580890498$\cdot 10^{-4}$ & -8.5206014973391348 \\
  Ar$^{16+}$ & 18 & 9.684480583887134$\cdot 10^{-3}$ & 2.364226371363055$\cdot 10^{-4}$ & -9.0207335455792636 \\
             \hline 
  K$^{17+}$  & 19 & 8.671093473064351$\cdot 10^{-3}$ & 1.894471957484780$\cdot 10^{-4}$ & -9.5208515720108625 \\
  Ca$^{18+}$ & 20 & 7.808833943501406$\cdot 10^{-3}$ & 1.535808715634216$\cdot 10^{-4}$ & -10.020957697600605 \\ 
  Sc$^{19+}$ & 21 & 7.069079045780357$\cdot 10^{-3}$ & 1.258148786675343$\cdot 10^{-4}$ & -10.521053636055015 \\ 
  Ti$^{20+}$ & 22 & 6.429674915299672$\cdot 10^{-3}$ & 1.040496858269455$\cdot 10^{-4}$ & -11.021140787100487 \\
  V$^{21+}$  & 23 & 5.873258366855471$\cdot 10^{-3}$ & 8.679410260860985$\cdot 10^{-5}$ & -11.521220305246443 \\
  Cr$^{22+}$ & 24 & 5.386065414197868$\cdot 10^{-3}$ & 7.297190978578113$\cdot 10^{-5}$ & -12.021293151234381 \\
              \hline
  Mn$^{23+}$ & 25 & 4.957071712852294$\cdot 10^{-3}$ & 6.179497295229088$\cdot 10^{-5}$ & -12.521360131054423 \\
  Fe$^{24+}$ & 26 & 4.577363304001367$\cdot 10^{-3}$ & 5.267838395821848$\cdot 10^{-5}$ & -13.021421925898852 \\
  Co$^{25+}$ & 27 & 4.239669745906067$\cdot 10^{-3}$ & 4.518271622832327$\cdot 10^{-5}$ & -13.521479115417290 \\
  Ni$^{26+}$ & 28 & 3.938013446919951$\cdot 10^{-3}$ & 3.897410055792940$\cdot 10^{-5}$ & -14.021532195958377 \\ 
     \hline\hline
  \end{tabular}}
  \end{center}
  \end{table}
\newpage
 \begin{table}[tbp]
   \caption{The $\langle r^{-1}_{ee} \rangle, \langle -\frac12 \nabla^2_e \rangle$ and $\langle \delta_{321} \rangle$ 
            expectation values determined for the different two-electron atoms/ions in their ground $1^1S-$states (in 
            atomic units). All nuclear masses are assumed to be infinite.}
     \begin{center}
     \scalebox{0.85}{%
     \begin{tabular}{| l | l | l | l | l |}
      \hline\hline
  ion/atom & $Q$ & $\langle r^{-1}_{ee} \rangle$ & $\langle -\frac12 \nabla^2_e \rangle$ & $\langle \delta_{321} \rangle$ \\
     \hline\hline              
   H$^{-}$     & 1 & 0.3110215022143000515 & 0.26387550827218860 & 0.00512686786 \\    
   He          & 2 & 0.9458184487999231936 & 1.45186218851705980 & 1.86179097423 \\ 
   Li$^{+}$    & 3 & 1.5677195591374732721 & 3.63995670633465298 & 33.3211589496 \\
   Be$^{2+}$   & 4 & 2.1908707739064748156 & 6.82778311921179335 & 231.038802708 \\
   B$^{3+}$    & 5 & 2.8146960470665581690 & 11.0154857901213908 & 996.009538154 \\
   C$^{4+}$    & 6 & 3.4388907009922270508 & 16.2031233009492652 & 3221.84957561 \\
             \hline 
   N$^{5+}$   & 7  & 4.0633059371614058208 & 22.3907225743863523 & 8596.07234611 \\
   O$^{6+}$   & 8  & 4.6878626272933908613 & 29.5782975613789628 & 19974.3019384 \\
   F$^{7+}$   & 9  & 5.3125152146355572467 & 37.7658561819797455 & 41827.4545255 \\
  Ne$^{8+}$   & 10 & 5.9372357244481370617 & 46.9534032575187747 & 80761.8501364 \\
  Na$^{9+}$   & 11 & 6.5620060637014121725 & 57.1409418880363610 & 14611.2288400 \\
  Mg$^{10+}$  & 12 & 7.1868140253485781079 & 68.3284741563234650 & 250608.071842 \\
             \hline 
  Al$^{11+}$ & 13 &  7.8116510805680880023 & 80.5160015130291799 & 411111.980219 \\
  Si$^{12+}$ & 14 &  8.4365110930811231600 & 93.7035249993314632 & 49432.1960573 \\
  P$^{13+}$  & 15 &  9.0613895365056378705 & 107.891045381768580 & 993207.180907 \\
  S$^{14+}$  & 16 &  9.6862829995969480726 & 123.078563237127370 & 1476863.50256 \\
  Cl$^{15+}$ & 17 & 10.311188863149912371 & 139.2660790077000478 & 2142646.61319 \\
  Ar$^{16+}$ & 18 & 10.936105082961410952 & 156.4535930383055744 & 3041724.57849 \\
             \hline 
  K$^{17+}$  & 19 & 11.561030040385195148 & 174.6411056017265835 & 4235364.75740 \\
  Ca$^{18+}$ & 20 & 12.185962437154670069 & 193.8286169165792781 & 5796183.43295 \\ 
  Sc$^{19+}$ & 21 & 12.810901219907308976 & 214.0161271601173455 & 7809468.39351 \\ 
  Ti$^{20+}$ & 22 & 13.435845525071932709 & 235.2036364775691919 & 10374574.4651 \\
  V$^{21+}$  & 23 & 14.060794637989284133 & 257.3911449890558868 & 13606391.9942 \\
  Cr$^{22+}$ & 24 & 14.685747962156786948 & 280.5786527947906359 & 17636888.2813 \\
              \hline
  Mn$^{23+}$ & 25 & 15.310704995789691498 & 304.7661599790378727 & 22616721.9649 \\
  Fe$^{24+}$ & 26 & 15.935665313746399507 & 329.9536666131639021 & 28716930.3559 \\
  Co$^{25+}$ & 27 & 16.560628553438948983 & 356.1411727580132750 & 36130689.7235 \\
  Ni$^{26+}$ & 28 & 17.185594403740284234 & 383.3286784657785490 & 45075148.5302 \\
     \hline\hline
  \end{tabular}}
  \end{center}
  \end{table}
\newpage
 \begin{table}[tbp]
   \caption{The coefficients of the $Q^{-1}$ expansion determined for the $\langle r^{-1}_{ee} \rangle$ expectation values
            determiend for all 28 ions presented in Table IV (in atomic units). $N$ is the total number of terms used in 
            the $Q^{-1}$ expansion.}
     \begin{center}
     \scalebox{0.75}{%
     \begin{tabular}{| l | l | l | l | l |}
      \hline\hline
  cefficient & $N = 10$  & $N = 10$ & $N = 12$  & $N = 14$ \\
        \hline
  $A_2$     &   -0.494831796599196E-08 &  0.000000000000000E+00 &  0.000000000000000E+00 &  0.000000000000000E+00 \\
  $A_1$     &    0.625000467182303E+00 &  0.624999946424256E+00 &  0.625001683616873E+00 &  0.625003246752918E+00 \\
  $A_0$     &   -0.315350485009514E+00 & -0.315327955933086E+00 & -0.315514341765753E+00 & -0.315701190374814E+00 \\
        \hline
  $B_1$     &    0.264528972963422E-01 &  0.259297624553894E-01 &  0.343629291054843E-01 &  0.439685640740424E-01 \\
  $B_2$     &   -0.764173015283037E-02 & -0.468906451156542E-03 & -0.212091717363386E+00 & -0.492221452096754E+00 \\
  $B_3$     &    0.217453679005328E-01 & -0.386848769404071E-01 &  0.322514909365720E+01 &  0.837712175158511E+01 \\
  $B_4$     &   -0.104596000674926E+00 &  0.211498434740675E+00 & -0.322114782408164E+02 & -0.950787058977685E+02 \\
  $B_5$     &    0.203688490986952E+00 & -0.807975095629530E+00 &  0.210536612541492E+03 &  0.731912378303186E+03 \\
         \hline
  $B_6$     &   -0.206581093655412E+00 &  0.168265813518147E+01 & -0.900343198369528E+03 & -0.385843989156749E+04 \\
  $B_7$     &    0.683035978612077E-01 & -0.177678156292532E+01 &  0.246014385082417E+04 &  0.138443324014201E+05 \\
  $B_8$     &                          &  0.705173625864056E+00 & -0.406809330705356E+04 & -0.330327568745938E+05 \\
  $B_9$     &                          &                        &  0.362963884774198E+04 &  0.499600243794525E+05 \\
  $B_{10}$  &                          &                        & -0.130271721358420E+04 & -0.437013456795518E+05 \\
          \hline
  $B_{11}$  &                          &                        &                        &  0.182548118898741E+05 \\
  $B_{12}$  &                          &                        &                        & -0.211138704685216E+04 \\
     \hline\hline
  \end{tabular}}
  \end{center}
  \end{table}
\newpage
 \begin{table}[tbp]
   \caption{The lowest order QED-corrections $\Delta E^{QED}_{\infty}$, Eq.(\ref{eqf8}), and numerical values of $\ln K_0, \langle \frac{1}{r^{3}_{ee}} \rangle, 
            \langle \frac{1}{r^{3}_{ee}} \rangle$ in atomic units. In this Table the notations $\Delta E^{QED}_{2;3}$ stands for the sum of the last two terms 
            from Eq.(\ref{eqf9}) multiplied by a factor $\frac{M}{m_e} = M$ (nuclear mass). All nuclear masses are assumed to be infinite.}
     \begin{center}
     \scalebox{0.75}{%
     \begin{tabular}{| l | l | l | l | l | l | l |}
      \hline\hline
  ion/atom & $Q$ & $\ln K_0$ & $\langle \frac{1}{r^{3}_{ee}} \rangle$ & $\langle \frac{1}{r^{3}_{ee}} \rangle$ &  $\Delta E^{QED}_{\infty}$ & $\Delta E^{QED}_{2;3}$ \\
     \hline\hline              
   H$^{-}$     & 1 & 2.9930441530  &  0.987145110780E-01 & -0.1367762464713E+01 & 0.12485643657E-05 &  0.555264710871E-07 \\    
   He          & 2 & 4.37016021981 &  0.989273545062E+00 & -0.3092590081520E+02 & 0.22261831904E-04 & -0.135298475377E-04 \\ 
   Li$^{+}$    & 3 & 5.179849129   &  0.178846906701E+00 & -0.1526042743967E+03 & 0.11024751764E-03 & -0.197385810727E-03 \\   
   Be$^{2+}$   & 4 & 5.755091813   & -0.758869771350E+01 & -0.4461493461302E+03 & 0.33035865562E-03 & -0.116481400438E-02 \\
   B$^{3+}$    & 5 & 6.201467201   & -0.289702881864E+02 & -0.1000390699060E+04 & 0.75812296982E-03 & -0.440940827548E-02 \\
   C$^{4+}$    & 6 & 6.566235883   & -0.715558284877E+02 & -0.1910673007667E+04 & 0.14744207461E-02 & -0.127893001871E-01 \\
             \hline 
   N$^{5+}$   & 7  & 6.874659414   & -0.143641108746E+03 & -0.3277442371480E+04 & 0.25617115664E-02 & -0.310542023859E-01 \\
   O$^{6+}$   & 8  & 7.141831512   & -0.254091119110E+03 & -0.5205342434111E+04 & 0.41010004202E-02 & -0.664058639717E-01 \\
   F$^{7+}$   & 9  & 7.377493052   & -0.412248230953E+03 & -0.7802584739240E+04 & 0.61692976733E-02 & -0.129086002178E+00 \\
  Ne$^{8+}$   & 10 & 7.588297026   & -0.627866111317E+03 & -0.1118048429795E+05 & 0.88374333231E-02 & -0.232987591610E+00 \\
  Na$^{9+}$   & 11 & 7.778989583   & -0.911059814059E+03 & -0.1545310262982E+05 & 0.12168138196E-01 & -0.396286477650E+00 \\
  Mg$^{10+}$  & 12 & 7.953075479   & -0.127226672979E+04 & -0.2073696471413E+05 & 0.16214333661E-01 & -0.642090983605E+00 \\
             \hline 
  Al$^{11+}$ & 13 & 8.1132164336   & -0.172221518425E+04 & -0.3466213164326E+06 & 0.21017588805E-01 & -0.109131248829E+01 \\
  Si$^{12+}$ & 14 & 8.2614815231   & -0.227189863118E+04 & -0.3481549672596E+05 & 0.26606715083E-01 & -0.150232170036E+01 \\
  P$^{13+}$  & 15 & 8.3995110079   & -0.293255406291E+04 & -0.4385365145496E+05 & 0.32996475895E-01 & -0.219369069970E+01 \\
  S$^{14+}$  & 16 & 8.5286271993   & -0.371564368022E+04 & -0.5438972182745E+05 & 0.40186393648E-01 & -0.312285083765E+01 \\
  Cl$^{15+}$ & 17 & 8.6499116644   & -0.463283913392E+04 & -0.6654976314062E+05 & 0.48159640601E-01 & -0.434783445856E+01 \\
  Ar$^{16+}$ & 18 & 8.7642603335   & -0.569600783126E+04 & -0.8046135394383E+05 & 0.56882002451E-01 & -0.593579835268E+01 \\
             \hline 
  K$^{17+}$  & 19 & 8.8724236914   & -0.691720092580E+04 & -0.9625350501886E+05 & 0.66300905715E-01 & -0.796376203590E+01 \\
  Ca$^{18+}$ & 20 & 8.9750366476   & -0.830864269774E+04 & -0.1140565786357E+06 & 0.76344501477E-01 & -0.105193554360E+02 \\
  Sc$^{19+}$ & 21 & 9.0726411126   & -0.988272109625E+04 & -0.1340022164342E+06 & 0.86920799336E-01 & -0.137015754571E+02 \\
  Ti$^{20+}$ & 22 & 9.1657033175   & -0.116519792627E+05 & -0.1562232746201E+06 & 0.97916846311E-01 & -0.176215509500E+02 \\
  V$^{21+}$  & 23 & 9.2546272805   & -0.136291078896E+05 & -0.1808537654137E+06 & 0.10919794629E+00 & -0.224033156641E+02 \\
  Cr$^{22+}$ & 24 & 9.3397654052   & -0.158269382973E+05 & -0.2080288038885E+06 & 0.12060691617E+00 & -0.281845887975E+02 \\
              \hline
  Mn$^{23+}$ & 25 & 9.4214269146   & -0.182584361318E+05 & -0.2378845594865E+06 & 0.13196337542E+00 & -0.351175627958E+02 \\
  Fe$^{24+}$ & 26 & 9.4998846295   & -0.209366956035E+05 & -0.2705582116209E+06 & 0.14306306617E+00 & -0.433696980840E+02 \\
  Co$^{25+}$ & 27 & 9.5753804665   & -0.238749342007E+05 & -0.3061879088677E+06 & 0.15367720134E+00 & -0.531245244377E+02 \\
  Ni$^{26+}$ & 28 & 9.6481299371   & -0.270864878203E+05 & -0.3449127313289E+06 & 0.16355183851E+00 & -0.645824487293E+02 \\ 
     \hline\hline
  \end{tabular}}
  \end{center}
  \end{table}
\end{document}